\documentclass[pra]{revtex4-2}

\usepackage{graphicx}
\usepackage{color}
\begin{document}

\title{Sequential propagation of a single photon through five \\ measurement contexts in a three-path interferometer}

\author{Holger F. Hofmann}
\email{hofmann@hiroshima-u.ac.jp}
\affiliation{
Graduate School of Advanced Science and Engineering, Hiroshima University,
Kagamiyama 1-3-1, Higashi Hiroshima 739-8530, Japan
}

\begin{abstract}
Quantum contextuality describes scenarios in which it is impossible to explain the experimental evidence in terms of a measurement independent reality. Here, I introduce a three-path interferometer in which all five contexts needed for a demonstration of contextuality are realized in sequence. It is then possible to observe a paradoxical situation where the paths connecting input ports to their corresponding output ports appear to be blocked by destructive interference. It is shown that the conditional currents observed in weak measurements provide a consistent explanation of the paradox, indicating that weak values might help to bridge the gap between wavelike propagation effects and local particle detection.
\end{abstract}

\maketitle

%%%%%%%%%%%%%%%%%%%%%%%%%%  body  %%%%%%%%%%%%%%%%%%%%%%%%%%

\section{Introduction}

Over the past decades, research into the foundations of quantum mechanics has resulted in a wide variety of new technologies based on the precise control of quantum states and their experimental characterization. For most researchers in the field, the paradoxical nature of quantum statistics has been a core motivation of their research. Yet all of the technological advances have only highlighted the counter-intuitive nature of quantum mechanics.  It may well be that the added levels of complexity are distracting us from the real progress that has been achieved by the thorough study of basic concepts such as the measurement postulate (also known as Born's rule) and the closely related problems of superpositions and quantum interference effects. Perhaps the most striking examples of the paradoxical nature of quantum statistics are the demonstrations of quantum contextuality, where it is shown that the statistics observed in different measurement contexts cannot be reconciled with any context independent reality of the measurement outcomes \cite{Koc67}. The most compact formulation of this problem was originally given by Hardy for entangled systems \cite{Har92,Har93} and was later generalized to include the most basic case of quantum contextuality involving the relations between measurement outcomes in a three-dimensional Hilbert space \cite{Kly08,Cab13}. 

It is perhaps not surprising that the basic problem of quantum contextuality in three-level systems has been discovered more than once by researchers with different backgrounds and ideas. It seems to have made its earliest appearance as the three-box paradox, which is more commonly associated with weak values \cite{Aha91,Res04}, even though it was already related to a Hardy-like proof of non-contextuality by Clifton in 1993 \cite{Cli93,Lei05}. It was only recognized later that it can also be obtained by systematically reducing the original proof of contextuality by Kochen and Specker to the minimal number of necessary measurement outcomes \cite{Kly08,Cab13}. As these compact proofs of contextuality show, a total of five measurement contexts are needed to make the argument complete. In principle, the experimental confirmation of this basic paradox can thus be achieved by performing separate measurements on a given input state, combining the data of the different measurements to show that the probabilities exceed non-contextual limits \cite{Ahr13}. However, such separate measurements do not tell us much about the relation between the different measurement contexts, since each measurement requires a completely different setting of the experimental apparatus. Indeed, one of the fundamental questions investigated using a three-path interferometer was the possibility of changing contexts without any effect on the measurement probability of an outcome shared by two different contexts \cite{Mar14}. It is also worth noting that a generalized notion of contextuality introduced by Spekkens \cite{Spe05} has recently been linked to interferences in a Mach-Zehnder interferometer \cite{Pan20,Wag22}, demonstrating how closely single particle interference effects are linked to the fundamental notion of contextuality in quantum mechanics \cite{Dre12,Pus14,Dre15,Kun19,Wag23}. It may therefore be useful to design a more elaborate setup that combines different measurement contexts with each other in a sequence of interferences between three paths to directly implement a Kochen-Specker contextuality paradox \cite{Kly08,Cab13}. In this paper, I introduce a three-path interferometer in which all five measurement contexts needed for the demonstration of Kochen-Specker contextuality are physically implemented by a sequence of beam splitters. The interferometer is constructed in such a way that the input context and the output context are identical, so that the output port at which a photon is detected corresponds to a detection of the input port. The problem of contextuality is then directly related to the problem of reconstructing the path taken by a single particle through the interferometer as they propagate from a well-defined input port to the corresponding output port. 

In the three-path interferometer introduced in this paper, the contextuality paradox describes a situation where the only available paths between an input port and its corresponding output port are apparently empty as quantum interferences reduce the probability of finding a photon in them to zero. Effectively, destructive interferences seem to block the paths that are necessary to explain the propagation of the photon through the interferometer. Interestingly, this problem also exists in classical wave interference, where it suggests inconsistencies between the flow of energy through the interferometer and the transmission of the amplitudes. The main difference between this classical problem of wave propagation and quantum contextuality originates from the individual detection of particles at the output of the interferometer that requires a definition of conditional paths or currents associated with only one output port for the explanation of particle propagation. Here, it is shown that a consistent description of conditional currents can be obtained from the weak values of path projectors \cite{Lun11,Lun12,Hof12,Hof15,Bud23}. The algebra of weak values ensures the continuity of conditional currents and the faithful assignment of the correct input port to each output port for all possible input states. Quantum contextuality is associated with negative values for the conditional currents that arise from quantum coherences between the detected output port and alternative output ports, where each coherence introduces its own specific bias between the paths. It is possible to identify a specific set of negative conditional currents that is associated with the observation of contextuality. Weak values and their associated conditional currents thus provide a natural link between the wavelike propagation of quantum particles and their localized detection in only one of the output paths, highlighting the difficulties of associating quantum state components and their amplitudes with measurement outcomes when these measurements are not actually performed \cite{Auf19,Han22,Mat23}.

\section{Contextuality in a three-path interferometer}

\begin{figure}[ht]
\vspace{-1cm}
\begin{picture}(360,200)
%%\put(0,0){\framebox(360,200){}}
\put(0,0){\makebox(360,200){\vspace{-3cm}
\scalebox{0.8}[0.8]{
\includegraphics{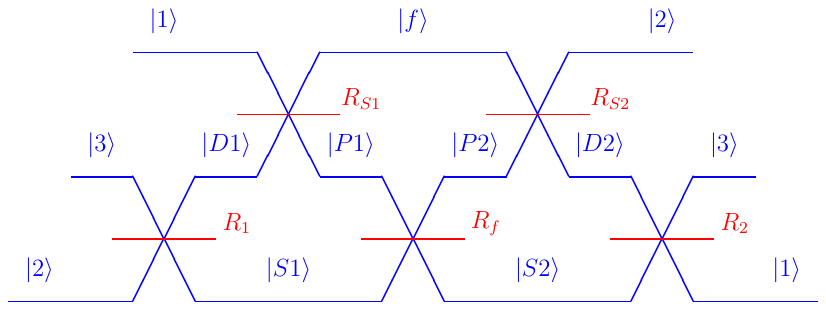}}}}
\end{picture}
%%\vspace{-5cm}
\caption{\label{paths}
Schematic representation of the three-path interferometer, where the paths represent five different measurement contexts being transformed into each other by beam splitters of reflectivities $R_i$. The input ports and the output ports represent the same context, but the order has been reversed. 
}
\end{figure}

The most compact formulation of a Kochen-Specker contextuality paradox is obtained when five different contexts of the three-dimensional Hilbert space are linked by five shared measurement outcomes \cite{Cab13,Ji23}. If two contexts share one measurement outcome, the change from one context to the other is represented by the transformation of the remaining two outcomes. In a single particle interference experiment, this transformation can be realized by a single beam splitter. It is therefore possible to construct an interferometer in which all five contexts appear as parallel paths by sequentially interfering pairs of paths at five different beam splitters. If we start with the paths $\{1,2,3\}$, the first beam splitter interferes path $2$ and path $3$ with a reflectivity of $R_1$, where reflection of $2$ connects to a new output path $S1$ and reflection of $3$ connects to a new output path $D1$. The new context $\{1,S1,D1\}$ can then be transformed at a second beam splitter that interferes the paths $1$ and $D1$ with a reflectivity of $R_{S1}$, where reflection of $1$ connects to a new output path $f$ and reflection of $D1$ connects to a new output path $P1$. The third context $\{f,S1,P1\}$ is transformed by interference between $S1$ and $P1$ at a reflectivity of $R_f$, where reflection of $S1$ connects to $S2$ and reflection of $P1$ connects to $P2$. As the choice of labels shows, the role of $1$ and $2$ have now been exchanged and the remaining two beam splitters of reflectivity $R_{S2}$ and $R_2$ transform the context $\{f,S2,P2\}$ first into $\{2,S2,D2\}$ and then back to the first context, $\{1,2,3\}$. The complete three path interferometer is shown in Fig. \ref{paths}. Note that it is possible to distinguish between five outer paths $1, S1, f, S2$ and $2$ that appear in two different contexts each, and five inner paths $3, D1, P1, P2$ and $D2$ that appear in only one context each. 

Each of the paths represent measurement outcomes that can be represented by Hilbert space vectors. The unitary transformations represented by the beam splitters are defined by inner products between the Hilbert space vectors. Different phase conventions are possible, but the symmetry of the setup strongly suggests that the reflection of the inner paths should obtain a negative sign. For example, the unitary transformation of the central beam splitter with reflectivity $R_f$ can be expressed by
\begin{eqnarray}
\mid S2 \rangle &=& \sqrt{R_f} \mid S1 \rangle + \sqrt{1-R_f} \mid P1 \rangle,
\nonumber \\
\mid P2 \rangle &=& \sqrt{1-R_f} \mid S1 \rangle- \sqrt{R_f} \mid P1 \rangle.
\end{eqnarray}
The reflectivities themselves are not independent of each other, but need to be chosen in a specific manner to ensure that the output states $\{\mid 1 \rangle, \mid 2 \rangle, \mid 3 \rangle\}$ are equal to the input states $\{\mid 1 \rangle, \mid 2 \rangle, \mid 3 \rangle\}$. This cyclic condition can only be satisfied if the connection of two outer paths by reflection at the beam splitter connecting them directly is equal to the product of transmissions through the two beam splitters that connect the same outer paths via an inner path. For the outer paths $S1$ and $S2$, this inner path is path $3$, and the relation between the reflectivities is
\begin{equation}
R_f = (1-R_1)(1-R_2).
\end{equation}
The reflectivity of the central beam splitter can therefore be derived from the reflectivities of the first and the last beam splitter. The remaining two reflectivities can be defined likewise, using the outer path combinations $(1,S2)$ and $(2,S1)$, 
\begin{eqnarray}
1-R_{S1} &=& \frac{R_2}{1-R_f},
\nonumber \\
1-R_{S2} &=& \frac{R_1}{1-R_f}.
\end{eqnarray}
A contextual three-path interferometer can be constructed for any pair of values $(R_1,R_2)$ other than zero or one. The scenario that maximizes the probability of the paradoxical outcome is obtained for $R_1=R_2=1/2$, with $R_f=1/4$ and $R_{S1}=R_{S2}=1/3$ \cite{Cab13,Ji23}. In the following, this selection of reflectivities will be used for all of the examples. However, it should be remembered that the general analysis applies to all possible choices of the reflectivities $R_i$. For example, a completely symmetric solution with equal reflectivities for all five beam splitters can be found by solving the relation $R=(1-R)^2$, resulting in a reflectivity of $(3-\sqrt{5})/2=0.382$. Clearly, the case of $R_1=R_2=1/2$ is more convenient for any numerical calculations, but the results derived in the following are also valid in the fully symmetric case and in all other cases that can be constructed by varying $R_1$ and $R_2$. 

\begin{figure}[ht]
\vspace{-1cm}
\begin{picture}(360,200)
%%\put(0,0){\framebox(360,200){}}
\put(0,0){\makebox(360,200){\vspace{-3cm}
\scalebox{0.8}[0.8]{
\includegraphics{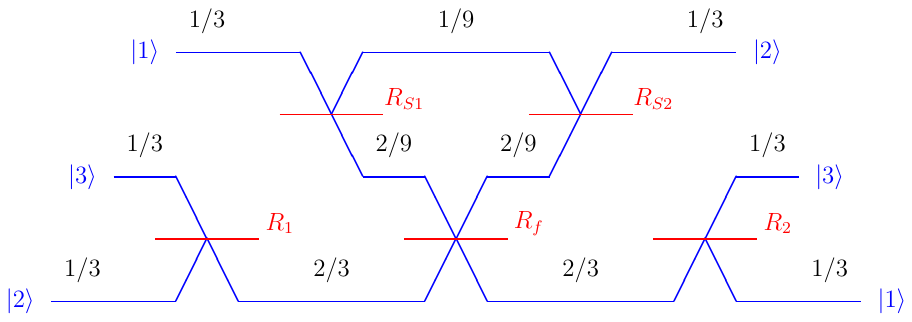}}}}
\end{picture}
%%\vspace{-5cm}
\caption{\label{paradox}
Illustration of contextuality for the input state $\mid N_f \rangle$ with $P(D1)=0$ and $P(D2)=0$. The numbers attached to each path give the probabilities of finding a photon in this path for reflectivities of $R_1=R_2=1/2$, $R_{S1}=R_{S2}=1/3$, and $R_f=1/4$. The fact that no photons are found in $D1$ or $D2$ is illustrated by the removal of the corresponding paths. If there is no probability current in $D1$ and $D2$, the probability in $f$ must originate in $1$ and end up in $2$. This is the Hardy-like contextuality paradox associated with the state $\mid N_f \rangle$.
}
\end{figure}

The construction of a three-path interferometer where five different contexts are represented by parallel paths perfectly maps the corresponding contextuality paradox to a basic interference problem. For any given input state, photon detection could be performed in any of the paths, providing a complete set of measurement results for all possible contexts. If the insertion of detectors is too difficult, photon polarization could be used to mark one path by a polarization flip. The detection probability for that path can then be obtained from the polarization statistics of the output photons. In general, a given input state is characterized by a specific distribution of path probabilities within the interferometer. 
%%%---redundant phrase cut, response to 6. of 2nd reviewer
In the Hardy-like paradox associated with the five contexts realized by the interferometer, the initial state is chosen so that $P(D1)=0$ and $P(D2)=0$. Since the paths $D1$ and $D2$ are empty, the only remaining path from input $1$ to output $1$ runs through $P1$ and $S2$, the only remaining path from input $2$ to output $2$ runs through $S1$ and $P2$, and the only remaining path from input $3$ to output $3$ runs through $S1$ and $S2$. The exclusion of $D1$ and $D2$ implied by $P(D1)=0$ and $P(D2)=0$ leaves no room for any path running through $f$, since such a path would have to run from an input of $1$ to an output of $2$. Fig. \ref{paradox} illustrates this problem by indicating the only available path through $f$ when both $D1$ and $D2$ are excluded. However, the quantum state defined by $P(D1)=0$ and $P(D2)=0$ has a non-vanishing probability of $P(f)>0$, indicating that some photons can be found in $f$. This raises the question of how the photons propagate through the interferometer when they are in the state defined by $P(D1)=0$ and $P(D2)=0$.

Before discussing the propagation of photons through the interferometer in more detail, it may be useful to provide some additional details on the quantum state defined by $P(D1)=0$ and $P(D2)=0$. Since the characteristic properties of this state suppress the probability of finding the photon in the path $f$, the state will be labeled $\mid N_f \rangle$. In the $\{1, S1,D1\}$-context, this state is given by the superposition
\begin{equation}
\mid N_f \rangle = \mid 1 \rangle\langle 1 \mid N_f \rangle + \mid S1 \rangle\langle S1 \mid N_f \rangle.
\end{equation}
The inner product of $\mid D2 \rangle$ and $\mid N_f \rangle$ is zero, and the inner products of $\mid D2 \rangle$ with the components $\mid 1 \rangle$ and $\mid S1 \rangle$ can be expressed using the reflectivities $R_1$ and $R_2$,
\begin{eqnarray}
\langle D2 \mid 1 \rangle &=& \sqrt{1-R_2},
\nonumber \\
\langle D2 \mid S1 \rangle &=& - \sqrt{R_2 (1-R_1)}.
\end{eqnarray}
These relations define the probability $P(1|N_f)=|\langle 1 \mid N_f \rangle|^2$ of finding the photon in path $1$ when the state $N_f$ is defined by $P(D1|N_f)=0$ and $P(D2|N_f)=0$. The result only depends on the reflectivities $R_1$ and $R_2$ and reads
\begin{equation}
|\langle 1 \mid N_f \rangle|^2 = \frac{R_2(1-R_1)}{1-R_1 R_2}.
\end{equation} 
Likewise, expanding $\langle D1 \mid N_f \rangle=0$ in the $\{2, S2, D2\}$-basis gives
\begin{equation}
|\langle 2 \mid N_f \rangle|^2 = \frac{R_1(1-R_2)}{1-R_1 R_2}.
\end{equation} 
Since the sum of the outcome probabilities in the $\{1, 2, 3\}$-context is one, the probability of the outcome $3$ is given by
\begin{equation}
|\langle 3 \mid N_f \rangle|^2 = \frac{(1-R_1)(1-R_2)}{1-R_1 R_2}.
\end{equation} 
In the case of $R_1=R_2=1/2$, all three probabilities are equal to $1/3$ and $\mid N_f \rangle$ is an equal superposition of the states $\{ \mid 1 \rangle, \mid 2 \rangle, \mid 3 \rangle \}$. 

At the first beam splitter, $P(2|N_f)$ and $P(3|N_f)$ are transmitted into $P(S1|N_f)$, leaving path $D1$ empty. The probability in path $1$ is then split into path $f$ and path $P1$ according to the reflectivity $R_{S1}$ of the second beam splitter,
\begin{equation}
P(f|N_f)=R_{S1} P(1|N_f).
\end{equation}
The combination of an empty path $D1$ and a non-vanishing probability of finding the photon in path $1$ results in a non-vanishing probability of finding the photon in path $f$. The precise probability $P(f|N_f)=|\langle f \mid N_f \rangle|^2$ depends only on the reflectivities $R_1$ and $R_2$,
\begin{equation}
|\langle f \mid N_f \rangle|^2=\frac{R_1 R_2 (1-R_1)(1-R_2)}{(1-R_1 R_2)(1-(1-R_1)(1-R_2))}.
\end{equation}
It should be noted that this result corresponds to the general result for Hardy-like contextuality paradoxes recently derived from Hilbert space relations between different measurement contexts \cite{Ji23}. The maximal probability of $P(f|N_f)=1/9$ is obtained for $R_1=R_2=1/2$. Fig. \ref{paradox} shows the probabilities of finding a photon in state $\mid N_f \rangle$ in the different paths of the interferometer. The paths $D1$ and $D2$ have been removed to highlight the fact that they are empty. It seems as if the photon current flowing through path $f$ must originate from path $1$ and end up in path $2$, although the input paths and the output paths are clearly represented by the same Hilbert space basis $\{ \mid 1 \rangle, \mid 2 \rangle, \mid 3 \rangle \}$. If the Hilbert space representation of photon detection is correct, a photon detected in state $\mid 2 \rangle$ must have entered the interferometer in state $\mid 2 \rangle$ and a photon detected in state $\mid 1 \rangle$ must have entered in state $\mid 1 \rangle$. Coherent propagation does not allow any transfer of amplitudes between states that describe alternative measurement outcomes in the same context. Even in classical wave interference, the assumption that wave intensity was transferred from input mode $1$ to output mode $2$ seems to be a rather extreme violation of the obvious causality relations between the input amplitudes and the output amplitudes. Since it does not provide any clear indication where the intensity in output mode $2$ came from, wave propagation fails to resolve the propagation paradox associated with quantum contextuality. A more detailed characterization of the quantum statistics of three-path interference is needed to identify the precise role of quantum coherences between the paths in the distribution of measurement probabilities.

\section{Weak measurements of conditional currents}

%%---cut in response to reviewer 1

It is possible to explore the causality of propagation in more detail by using weak modulations of the amplitudes that do not cause any significant changes of the output probabilities, e.g. by applying small rotations of polarization \cite{Lun11,Lun12,Vai13,Han23}. Such a procedure is equivalent to a weak measurement of the path projector of the path in which the modulation is applied \cite{Hof12,Bud23} and can thus provide experimental evidence that the previously discussed link between weak values and generalized contextuality \cite{Dre12,Pus14,Dre15,Kun19,Wag23} also applies to Kochen-Specker contextuality paradoxes \cite{Hof15}. In the following, weak values will be used to identify the conditional currents that passed through a path $i$ before arriving at an output $o$.

The results of weak measurements are given by weak values $W(i|x,o)$, where $i$ represents the path in which the modulation is applied, $x$ represents the initial state, and $o$ represents the output path in which the photon is detected. In general, these weak values are given by
\begin{equation}
\label{eq:WV}
W(i|x,o) = \frac{\langle o \mid i \rangle \langle i \mid \hat{\rho}_x \mid o \rangle}{\langle o \mid \hat{\rho}_x \mid o \rangle}, 
\end{equation}
where $\hat{\rho}_x$ is the (possibly unknown) quantum state $x$ in the input ports. Each weak value $W(i|x,o)$ is a complex number that can be associated with the density matrix elements of the quantum state $\hat{\rho}_x$ in the output basis $\{\mid o \rangle\}$. The weak values $W(i|x,o)$ can thus be used to trace the propagation of amplitudes through the interferometer. To distinguish between state-dependent conditional properties and state-independent relations between weak values, it will be convenient to omit the indicator $x$ whenever a relation applies to all possible input states. In the following, $W(i|o)$ will be used to represent the weak values associated with an intermediate path $i$ and a final outcome $o$ in relations that apply to all possible input states. For example, all input states necessarily satisfy $W(2|1)=0$ and $W(1|1)=1$, corresponding to the expectation that photons entering in $\mid 1 \rangle$ must exit in $\mid 1 \rangle$, and photons entering in $\mid 2 \rangle$ never exit in $\mid 1 \rangle$. However, we can also confirm that there is no conditional current in $D1$ or in $D2$ if the initial state is $\mid N_f \rangle$,
\begin{eqnarray}
W(D1|N_f,2)&=&0,
\nonumber \\
W(D2|N_f,1)&=&0.
\end{eqnarray}
Here, the preselected condition $N_f$ is included to indicate that these are specific weak values for that particular initial state. 

\begin{figure}[ht]
\vspace{-1cm}
\begin{picture}(360,410)
%%\put(0,0){\framebox(360,410){}}
\put(0,0){\makebox(360,410){%%\vspace{-0.5cm}
\scalebox{0.8}[0.8]{
\includegraphics{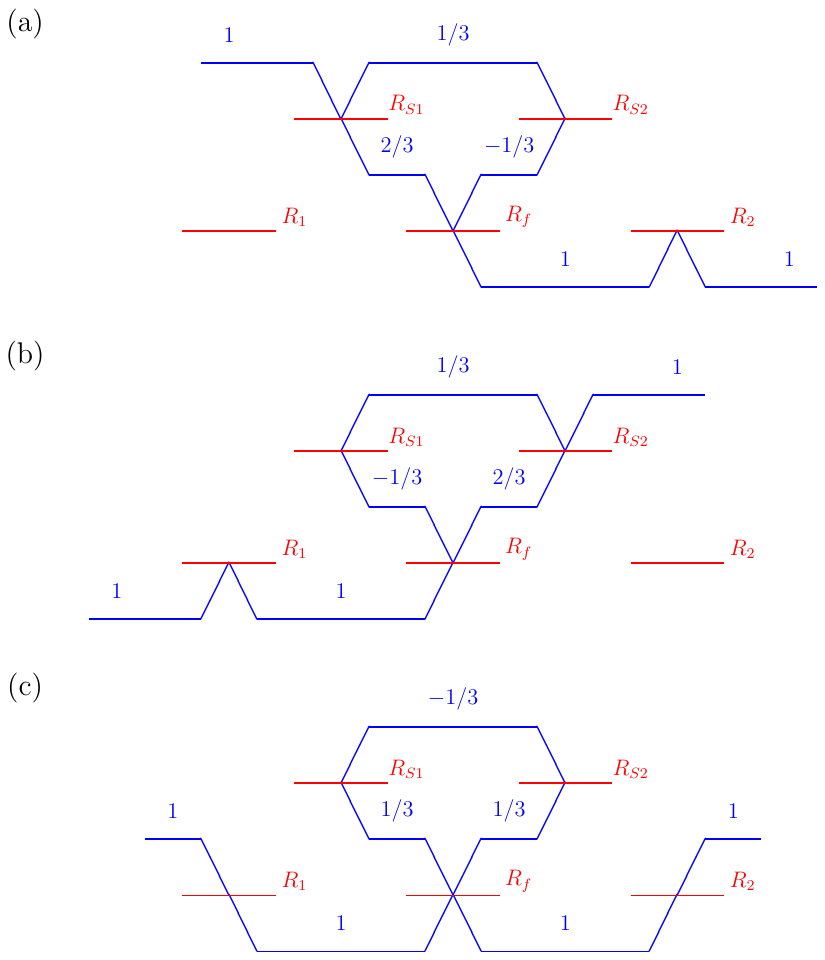}}}}
\end{picture}
%%\vspace{-5cm}
\caption{\label{Wgraph}
Weak values $W(i|N_f,o)$ of the paths $i$ conditioned by the detected outcomes $o$ for the input state $\mid N_f \rangle$. The values given are for reflectivities of $R_1=R_2=1/2$, $R_{S1}=R_{S2}=1/3$, and $R_f=1/4$. (a) shows the case of $o=1$, which occurs with a probability of $P(1)=1/3$. (b) shows the case of $o=2$, which occurs with a probability of $P(2)=1/3$. (c) shows the case of $o=3$, which occurs with a probability of $P(3)=1/3$. The probability of $P(f)=1/9$ is explained by the compensation of positive and negative conditional currents at the beam splitters marked $R_{S1}$ and $R_{S2}$ in the figures.
}
\end{figure}

The weak values of the path projectors in an interferometer are related to each other by continuity equations since the sum of the two projectors describing the input paths of a beam splitter must always be equal to the sum of the projectors on the output paths. This means that the weak values $W(f|N_f,1)$ and $W(f|N_f,2)$ can only be positive if the weak values $W(P2|N_f,1)$ and $W(P1|N_f,2)$ have the corresponding negative values,
\begin{eqnarray}
W(P1|N_f,2)&=& - W(f|N_f,2),
\nonumber \\
W(P2|N_f,1)&=& - W(f|N_f,1).
\end{eqnarray}
Likewise, the weak values $W(D1|N_f,3)$ and $W(D2|N_f,3)$ are both zero, so 
\begin{eqnarray}
W(P1|N_f,3)&=& - W(f|N_f,3),
\nonumber \\
W(P2|N_f,3)&=& - W(f|N_f,3).
\end{eqnarray}
Once the weak values of $f$ are determined, the conditional currents through the interferometer can be reconstructed. In general, the weak values are related to the total probability in $f$ by
\begin{equation}
\label{eq:fsum}
P(f) = W(f|1) P(1) + W(f|2) P(2) + W(f|3) P(3),
\end{equation}
where the state indicators $x$ have been omitted since the relation holds for all possible states.
Each contribution $W(f|o) P(o)$ corresponds to a quasi-probability assigning a joint statistical weight to $f$ and to $o$. These joint statistical weights satisfy continuity relations that can relate different outcomes to each other. For all states, the currents through $f$ satisfy
\begin{eqnarray}
\label{eq:Dcont}
W(f|D1) P(D1) &=& W(f|2) P(2) + W(f|3) P(3),
\nonumber \\
W(f|D2) P(D2) &=& W(f|1) P(1) + W(f|3) P(3).
\end{eqnarray}
For the state $\mid N_f \rangle$, the probabilities of $D1$ and $D2$ are zero, so the two contributions to the probability $P(f)$ on the right hand sides of Eq.(\ref{eq:Dcont}) must cancel. According to Eq.(\ref{eq:fsum}), this means that the weak values of $\mid N_f \rangle$ are related to each other by
\begin{eqnarray}
W(f|N_f,1) &=& P(f)/P(1),
\nonumber \\
W(f|N_f,2) &=& P(f)/P(2),
\nonumber \\
W(f|N_f,3) &=& - P(f)/P(3).
\end{eqnarray}
The weak values $W(i|N_f,o)$ describing the conditional currents through the three-path interferometer for the state $\mid N_f \rangle$ can thus be reconstructed completely from the probabilities $P(1)$, $P(2)$, $P(3)$ and $P(f)$ shown in Fig. \ref{paradox}. As shown in Fig. \ref{Wgraph}, the weak values then describe the paradoxical situation by assigning positive and negative conditional currents to $f$ and to $P1$ or $P2$, so that opposite currents through $f$ and $P1$ can originate at the second beam splitter ($R_{S1}$) without any input from $D1$, and opposite currents through $f$ and $P2$ can compensate each other at the fourth beam splitter ($R_{S2}$) without any transmission to $D2$. 

The discussion in this section shows that the weak values $W(o|N_f,i)$ provide a consistent explanation of the probability distribution of the input state $\mid N_f \rangle$ in the three-path interferometer without requiring any transfer of photons between the orthogonal states $\mid 1 \rangle$ and $\mid 2 \rangle$. The ``price'' for this explanation is the assignment of negative values to the conditional currents from $1$ to $1$ through $P2$ ($W(P2|N_f,1)$), from $2$ to $2$ through $P1$ ($W(P1|N_f,2)$, and from $3$ to $3$ through $f$ ($W(f|N_f,3)$). It is now possible to explore the physical origin of quantum contextuality by varying the input states, allowing for non-zero probabilities of $D1$ and $D2$. The non-contextual assumption that a photon passing through $f$ must either pass through $D1$ or through $D2$ on its way from any input path $o$ to the corresponding output path $o$ corresponds to the inequality
\begin{equation}
\label{eq:limit}
P(f) \leq P(D1) + P(D2).
\end{equation}
This inequality must always be satisfied if all of the conditional currents $W(i|o)$ have positive real parts. It is possible to divide the inequality into contributions associated with the postselection of one outcome $o$ each. These contributions can then be summarized using the continuity relations between conditional currents.

For all input states, the contribution associated with the outcome $o=1$ is given by
\begin{equation}
W(D2|1) - W(f|1) = W(P2|1),
\end{equation}
where a negative weak value $W(P2|1)$ contributes to the violation of Eq.(\ref{eq:limit}). Likewise, the contribution associated with $o=2$ is
\begin{equation}
W(D1|2) - W(f|2) = W(P1|2).
\end{equation}
It is more difficult to find a compact expression for the contribution associated with $o=3$. The contribution is best represented by a sum of three conditional currents,
\begin{equation}
W(D1|3)+W(D2|3)-W(f|3) = W(f|3)+W(P1|3)+W(P2|3).
\end{equation}
Combining these result, Eq.(\ref{eq:limit}) can be re-written as
\begin{equation}
\label{eq:negatives}
W(P2|1)P(1) + W(P1|2)P(2) + \left(W(f|3)+W(P1|3)+W(P2|3)\right) P(3) \geq 0.   
\end{equation}
This inequality provides a direct link between contextuality and negative conditional currents, identifying negative currents through $P1$, $P2$ and $f$ as the ones responsible for the experimentally observable violation of Eq.(\ref{eq:limit}).

The discussion above shows that the weak values of projectors on different paths provide a consistent description of conditional currents through the interferometer, where contextuality is described by negative currents. In all non-contextual models, conditional currents would have to represent the conditional probabilities of finding the photon in the respective path, prohibiting any negative values. Conversely, weak values can be interpreted as contextual measures of a deterministic distribution of individual particles as suggested in \cite{Hof21,Lem22,Hof23a}. It is then possible to allow negative currents in the context of a specific combination of pre- and postselection. Fig. \ref{Wgraph} shows the positive and negative current distribution of a single photon preselected in $\mid N_f \rangle$ and postselected in one of the three output ports. The algebra of weak values guarantees that the conditional currents are consistent with the standard formalism in which probabilities are derived from interferences between the amplitudes of different paths. The identification of the weak values $W(i|o)$ with conditional currents is therefore closer to the standard formalism of quantum mechanics than any speculation about potential hidden realities of localized photons. As I will explain below, this may even have implications for our understanding of classical wave interference, given that the mathematical formalism can be applied in an analogous manner.

\section{Coherences between output paths}

The weak values $W(i|o)$ provide a consistent characterization of conditional currents for all possible input states $\hat{\rho}_x$. The paradox of currents through $f$ seemingly running from input $1$ to output $2$ illustrated in Fig. \ref{paradox} is resolved by the assignment of negative conditional currents from $P1$ to $2$, from $P2$ to $1$, and from $f$ to $3$. The weak value formalism allows us to trace these negative conditional currents back to quantum coherences between the detected output and alternative outputs by expanding Eq.(\ref{eq:WV}) in the corresponding coherences between the post-selected outcome $o$ and the possible output paths $n=1,2,3$,
\begin{equation}
\label{eq:expand}
W(i|o) = \sum \frac{\langle o \mid i \rangle \langle i \mid n\rangle}{\langle o \mid \hat{\rho}_x \mid o \rangle} \langle n \mid \hat{\rho}_x \mid o \rangle.
\end{equation}
Note that the post-selection probability $P(o)=\langle o \mid \hat{\rho}_x \mid o \rangle$ appears in the denominator of every term. It is therefore convenient to multiply the weak value with the post-selection probability. The result is a linear combination of density matrix elements,
\begin{equation}
\label{eq:coherence}
 W(i|o) P(o) = \sum_n C(i|n,o) \langle n \mid \hat{\rho}_x \mid o \rangle.   
\end{equation}
This product of the weak value of a projector and the probability of post-selection corresponds to an element of a Kirkwood-Dirac (KD) distribution \cite{Lun12,Hof12,Hof15,Bud23}. In the present context, this means that the products of $W(i|o)$ and $P(o)$ provide an alternative description of the quantum state $\hat{\rho}_x$, where all measurement probabilities can be derived from linear relations between $W(i|o) P(o)$ and the density matrix elements $\langle n \mid \hat{\rho}_x \mid o \rangle$. 

To better understand the origin of negative conditional currents ($W(i|o)<0$), we can consider the manner in which coherences between the output paths re-distribute the conditional currents associated with the diagonal contributions given by
\begin{equation}
 C(i|o,o) \langle o \mid \hat{\rho}_x \mid o \rangle = |\langle i \mid o \rangle|^2 P(o).    
\end{equation}
These positive conditional currents can be observed directly, either by preparing $\mid o \rangle$ and detecting $\mid i \rangle$, or by preparing $\mid i \rangle$ and detecting $\mid o \rangle$. Since this corresponds to the scenarios described by textbook quantum mechanics, it is tempting to think between the paths $i$. For every complete orthogonal set of paths $\{\mid i \rangle\}$,
\begin{equation}
\sum_i C(i|n,o) = \langle o \mid n \rangle,    
\end{equation}
which is zero for $n \neq o$. In addition, Eq.(\ref{eq:expand}) shows that $C(i|n,0)$ is zero whenever $\mid i \rangle$ is orthogonal to either $\mid n \rangle$ or $\mid o \rangle$. For the coefficients $C(i|1,2)$, the only non-zero values are
\begin{equation}
\label{eq:C12}
C(P1|1,2)=C(P2|1,2)=-C(f|1,2).    
\end{equation}
Quantum coherence between the outcomes $1$ and $2$ re-distributed the conditional current of an outcome $2$ between path $f$ and the parallel path through $P1$ and $P2$. Since $C(i|n,o)=C(i|o,n)$ for real valued inner products between the states representing the paths, the same re-distribution applies to paths conditioned by an outcome of $1$. 

\begin{figure}[ht]
\vspace{-1cm}
\begin{picture}(360,280)
%%\put(0,0){\framebox(360,280){}}
\put(0,0){\makebox(360,280){\vspace{-7cm}
\scalebox{0.8}[0.8]{
\includegraphics{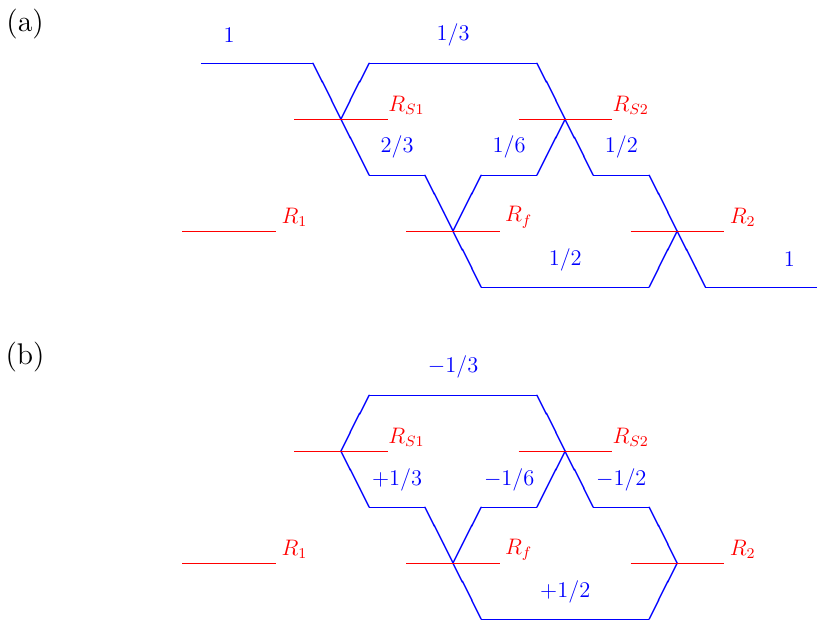}}}}
\end{picture}
%%\vspace{-5cm}
\caption{\label{bias}
Derivation of coefficients $C(i|3,1)$ from the difference between $W(i|1,1)=|\langle i \mid 1 \rangle|^2$ and the well-defined path from $1$ to $1$ via $S2$ given by $W(i|S2,1)$. (a) shows the values of $|\langle i \mid 1 \rangle|^2$ for $R_1=R_2=1/2$ and (b) shows the values of $C(i|3,1)$ describing the re-distribution of conditional currents caused by a quantum coherence between $3$ and $1$.
}
\end{figure}

The re-distribution of currents caused by coherences between $\mid 3 \rangle$ and $\mid 1 \rangle$ ($\mid 2 \rangle$) is a bit more complicated since it involves three different paths associated with $f$, $P2$ and $S2$ ($f$, $P1$ and $S1$). However. it is easy to see how the re-distribution works by considering the conditional currents of the input state $\mid S2 \rangle$ ($\mid S1 \rangle$). For this state and an outcome of $o=1$ ($o=2$), the conditional currents $W(i|S2,1)$ ($(W(i|S1,2)$) are one for $i=1$, $i=P1$ and $i=S2$ ($i=2$, $i=S2$ and $i=P1$), and zero for all other paths. This conditional current is obtained by re-distribution of the conditional currents $|\langle i \mid 1 \rangle|^2$ ($|\langle i \mid 1 \rangle|^2$) associated with the state $\mid 1 \rangle$ ($\mid 2 \rangle$). Based on this re-distribution, the coefficients $C(i|3,1)$ ($C(i|3,2)$) can be obtained directly from the differences of two sets of conditional currents, where $W(i|S2,1)$ ($(W(i|S1,2)$) represents the path of a particle and $|\langle i \mid 1 \rangle|^2$ ($|\langle i \mid 1 \rangle|^2$) represents the wavelike propagation associated with the identification of $\mid 1 \rangle$ ($\mid 2 \rangle$) with the input state. The coefficients can then be expressed as
\begin{eqnarray}
C(i|3,1) &=& \frac{\langle 1 \mid S2 \rangle}{\langle 3 \mid S2 \rangle} \left( W(i|S2,1) - |\langle i \mid 1 \rangle|^2\right), 
\nonumber \\
C(i|3,2) &=& \frac{\langle 2 \mid S1 \rangle}{\langle 3 \mid S1 \rangle} \left( W(i|S1,2) - |\langle i \mid 1 \rangle|^2\right). 
\end{eqnarray}
The results of this analysis for $C(i|3,1)$ with $R_1=R_2=1/2$ are shown in Fig. \ref{bias}. 

The state $\mid S2 \rangle$  ($\mid S1 \rangle$) satisfies the non-contextual expectation that the absence of photons in both $D1$ and $D2$ require an absence of photons in $f$. This expectation is satisfied because the suppression of conditional currents in $D2$ ($D1$) by the coherence between $\mid 3 \rangle$ and $\mid 1 \rangle$ ($\mid 2 \rangle$) also suppresses the conditional current in $f$. However, the addition of a coherence between $\mid 2 \rangle$ and $\mid 1 \rangle$ re-distributes conditional currents back into $f$, resulting in a violation of the non-contextual condition requiring $P(f)$ to be zero whenever $P(D1)=P(D2)=0$. The ideal violation of non-contextuality by the input state $\mid N_f \rangle$ is a direct consequence of the necessary coherence between $1$ and $2$ made necessary by the coherences between $1$ and $3$ and between $2$ and $3$ in the pure state limit. 

Eq.(\ref{eq:negatives}) identifies the negative conditional currents associated with contextuality. As Eq.(\ref{eq:C12}) shows, the coherence between $1$ and $2$ introduces a bias between $P1$ and $P2$ on one side, and $f$ on the other. Only $W(P1|2)$ and $W(P2|1)$ contribute to Eq.(\ref{eq:negatives}), so contextuality is observable when the coherence between one and two reduces the conditional currents in $P1$ and $P2$, while increasing the currents in $f$. This is confirmed by the negative values of $W(P1|2)$ and $W(P2|1)$ in Fig. \ref{Wgraph} (a) and (b). Fig. \ref{bias} shows that the coherence between $1$ and $3$ can reduce $W(f|3)$, $W(P2|3)$ and $W(P2|1)$. On the other hand, $W(P1|3)$ will increase as a result of this coherence. Likewise, the coherence between $1$ and $2$ can reduce $W(f|3)$, $W(P1|3)$ and $W(P1|2)$ while necessarily increasing $W(P2|3)$. Weak values thus reveal a fundamental relation between contextuality and the coherences between the output ports of the interferometer. Input state coherences between different potential output ports $o$ are an essential part of photon propagation through the interferometer, and these coherences are still part of the past of a particle when that particle has been detected at one of the three output ports. By describing the conditional effects of these coherences on the past of a particle, the weak values $W(i|o)$ bridge the gap between the wave-like propagation in the interferometer and the detection of individual particles in the output ports.

\section{Contextuality and wave-particle duality}

The three-path interferometer introduced above shows that the relation between five different measurement contexts at the heart of one of the most striking examples of quantum contextuality can be represented by the different paths inside the interferometer, where the change from one context to the next corresponds to interferences between two paths at one of the beam splitters. The specific sequence of interferences illustrates the relation between quantum contextuality and the interference of probability amplitudes, where paradoxes seem to arise because the presence of a non-zero probability in $f$ when quantum interference effects prevent any detection of photons in either $D1$ and $D2$ seems to suggest that photons enter the interferometer in path $1$ and exit in path $2$. However, the amplitudes entering in path $1$ are faithfully transmitted to path $1$, with no apparent connection between path $1$ and path $2$. Contextuality thus highlights a contradiction between the causality of particle propagation and the quantum mechanical formalism of interferences between paths.

In a classical wave formalism, the same situation arises, but its consequences are slightly different because we can differentiate between amplitudes and intensities. It is apparent that amplitudes are the carriers of causality in classical wave dynamics, where the transfer of energy is guided by the amplitudes. In the scenario shown in Fig. \ref{paradox}, the amplitudes from path $2$ and path $3$ reach both path $f$ and path $P2$, and the interference between path $f$ and path $P2$ ensures that only the amplitude from path $2$ exists in path $2$. However, it is impossible to assign separate amount of energy to these contributions. Specifically, it is not possible to separate the intensity in path $f$ into contributions from the amplitudes of $1$, $2$ and $3$. Contextuality paradoxes arise because particle detection requires us to associate the squares of amplitudes with probabilities of separate and distinct events. It is important to realize that it is not possible to reconcile the separation into distinct outcomes obtained in a specific measurement context with the separations obtained in other measurement contexts. It is therefore problematic to assume that wave propagation can be reconciled with particle propagation. The observation of a single localized particle in only one of the output ports can only be reconciled with the wave propagation of probability amplitudes if we accept that particle detection itself modifies the history of individual particles, so that the motion of a particle depends on future detection events. In particular, the localization of particles at the moment of detection conditions a past in which the particle was necessarily delocalized in a manner precisely defined by the corresponding weak values of projectors evaluating the presence of the particle in the paths \cite{Lem22,Hof23b}.   

The reason why weak values can provide a consistent description of contextual realities is that they cannot be determined before the measurements that select the context have been decided. Marking a path by a weak polarization rotation or a similar modulation of the amplitudes avoids the detection of a particle and provides no instantaneous information of the particle path. Nevertheless it is closely associated with the presence of a particle in the path and provides reliable information on the conditional currents through the interferometer. In terms of a generalized notion of contextuality, the weak values observed due to small polarization rotations cannot be reconciled with the statistics observed in discrete detection events \cite{Pus14}. With respect to wave-particle duality, the conditional currents defined by weak values indicate that the context of local photon detection at the output of the interferometer conditions a very specific pattern of delocalization of the same photon inside the interferometer. Only the final detection is discrete and local. The localization associated with the particle nature of photons is therefore contextual and does not apply to paths in which the photon is not detected. It may seem strange that the particle is automatically refocused on a single output port, but it needs to be remembered that quantum uncertainties prevent an observation of this process before the detection of the photon, indicating the absence of a measurement independent reality before the measurement has been performed.  

\section{Conclusions}

Quantum physics is difficult to understand because its simple mathematical structure gives rise to a wide range of counter-intuitive results. At the heart of this problem is the description of measurement, where a mathematical projection is used to attack a probability to each possible outcome. Since we usually identify measurement outcomes with measurement independent realities, the superposition of outcomes before the measurement is hard to interpret \cite{Auf19,Han22,Mat23}. Quantum contextuality shows that there is no way back to the classical idea of measurement independent realities. 
%%%%%%
In quantum optics, single particle superpositions are usually identified with classical waves, and the introduction of individual photons represents the actual challenge, both theoretically and experimentally. Optical interferometers are therefore ideally suited to investigate the role of quantum superpositions in the definition of measurement contexts. The relation between different measurement contexts then corresponds to the question of the path taken by an individual photon through the interferometer, and quantum contextuality shows that no consistent assignment of paths is possible. 
%%%%%%%
The interferometer introduced in this paper highlights the fundamental difference between the transmission of wave amplitudes and the transmission of particles along a specific path. In order to reconcile quantum interference effects with particle detection in the output of the interferometer, it is necessary to associate the effects of interferences between different output ports with only one of the two outcomes. Weak values serve this purpose, defining conditional currents that account for all quantum coherences between the different outcomes of photon detection in the output ports. 

In conclusion, the three-path interferometer introduced above strongly suggests that wave propagation cannot be reconciled with a realist notion of photons as localized particles moving along a well-defined path. The most elegant way to reconcile wave propagation with the detection of individual photons in the output ports is provided by weak values of the projection operators representing the potential detection of photons within the interferometer. These weak values define conditional currents that are consistent with the observable measurement probabilities. A realist interpretation of the conditional currents as conditional probabilities is prevented by the negative values of the conditional currents that are responsible for the observation of quantum contextuality. The inteferometric setup discussed here thus highlights the fundamental problems of wave-particle duality and presents us with a possible solution that may have important implications for the foundations of quantum mechanics.

\end{document}